\documentclass[10pt,letterpaper]{article}
\usepackage{opex3}
\usepackage{epsfig,amsmath,graphicx,amssymb}
\usepackage{graphicx}
\usepackage{dcolumn}
\usepackage{subfigure}
\usepackage{cite}
\usepackage{color}
\def\be{\begin{equation}}
\def\ee{\end{equation}}
\def\bea{\begin{eqnarray}}
\def\eea{\end{eqnarray}}
\def\bes{\begin{subequations}}
\def\ees{\end{subequations}}

\begin{document}

\title{Enhanced third-order and fifth-order Kerr nonlinearities in a cold atomic system via Rydberg-Rydberg interaction}

\author{Zhengyang Bai$^{1}$ and Guoxiang Huang$^{1,2}$}

\address{$^{1}$State Key Laboratory of Precision Spectroscopy and Department of Physics, East China Normal University, Shanghai 200062, China\\
         $^{2}$NYU-ECNU Joint Institute of Physics at NYU-Shanghai, Shanghai 200062, China}

\email{*gxhuang@phy.ecnu.edu.cn }

\begin{abstract}
We investigate the optical Kerr nonlinearities of an ensemble of cold Rydberg atoms under the condition of electromagnetically induced transparency (EIT). By using an approach beyond mean-field theory, we show that the system possesses not only enhanced third-order nonlinear optical susceptibility, but also giant fifth-order nonlinear optical susceptibility, which has a cubic dependence on atomic density. Our results demonstrate that both the third-order and the fifth-order nonlinear optical susceptibilities consist of two parts, contributed respectively by photon-atom interaction and Rydberg-Rydberg interaction. The Kerr nonlinearity induced by the Rydberg-Rydberg interaction plays a leading role at high atomic density. We find that the fifth-order nonlinear optical susceptibility in the Rydberg-EIT system may be five orders of magnitude larger than that obtained in traditional EIT systems. The results obtained  may have promising applications in light and quantum information processing and transmission at weak-light level.
\end{abstract}



\section{Intruduction}{\label{Sec1}}

The study of optical Kerr effect, i.e. nonlinear response of optical materials to applied light field, is one of  main topics in nonlinear optics because it is essential for the realization of most nonlinear optical processes~\cite{Boyd}. Optical Kerr effect has also found many new applications, including nonlinear and quantum controls of light fields, quantum nondemolition measurement, all-optical deterministic quantum logic, single-photonic switches and transistors, and so on~\cite{Nie,Chang}. However, Kerr effect is usually produced in passive optical media such as glass-based optical fibers, in which far-off resonance excitation schemes are employed to avoid serious optical absorption. As a result, the Kerr nonlinearity in passive optical media is weak and hence to obtain a significant Kerr effect a long propagation distance or a high light intensity is required.

In recent years, many efforts have focused on the study of electromagnetically induced transparency (EIT)~\cite{FIM,Khu}. Light propagation in EIT media possesses many striking features, including the suppression of optical absorption, the reduction of group velocity, and an enhancement of Kerr nonlinearity~\cite{FIM,Khu,Michinel0}, by which many important applications (e.g. quantum memory, highly efficient four-wave mixing, optical clocks, and slow-light solitons, etc.) are possible~\cite{FIM,Khu,Michinel0,Lvo,Santra,Zanon,Wu,Huang,Hang1,Chen}. However, the largest Kerr nonlinearity, obtained in conventional EIT media~\cite{Hau}, is still too small for nonlinear optics at single-photon level~\cite{Pri}.

Recently, much attention has been paid to the investigation of cold Rydberg gases~\cite{Pri,Gal,Saf}, i.e. highly excited atoms with very large principal quantum number. Due to their long lifetime and large electric dipole moment, Rydberg atoms have many practical applications~\cite{Pri1}. Especially, the strong and controllable atom-atom interaction in Rydberg gases (called Rydberg-Rydberg interaction for short~\cite{Saf,Pri1,Sevincli0}) brings many intriguing aspects that can be used to design quantum gates and simulate strongly correlated quantum many-body systems, etc~\cite{Pri,Gal,Saf,Pri1,Sevincli0,Henkel}.

Since the first experiment reported in 2007~\cite{Moha}, considerable achievements have been made on the EIT in cold Rydberg gases. Experimental~\cite{Mohapatra,pritchard1,pritchard2,Parigi,Hofmann,Maxwell2013} and theoretical~\cite{singer,Gorshkov,Sevincli,Ates,Gorshkov1,stan,Petrosyan,Yan} works
showed that EIT can be used not only for coherent optical detection of Rydberg atoms, but also for
obtaining giant Kerr nonlinearity~\cite{note00}. Different from conventional EIT media, the giant Kerr nonlinearity in Rydberg-EIT systems comes from the strong Rydberg-Rydberg interaction between atoms, which can be many orders of magnitude larger than those obtained before. These studies~\cite{Moha,Mohapatra,pritchard1,pritchard2,Parigi,Hofmann,Maxwell2013,singer,Sevincli,Ates,
Gorshkov,stan,Petrosyan,Yan,Gorshkov1,stan} opened a new and important avenue for the nonlinear optics
at single-photon level~\cite{Chang,Pey,First}.

However, all studies up to now on the Kerr nonlinearity in Rydberg-EIT systems are limited to
the third-order one. Because of the requirement of many applications in quantum and nonlinear optics,
such as highly efficient six-wave mixing~\cite{Saltiel,Vaicaitis}, three-photon phase gates~\cite{Zubairy,li},
multi-photon entangled states and Schr\"{o}dinger cat states of light~\cite{Auffeves,Haroche},
and stabilization of spatial optical solitons~\cite{Michinel0,Zhang},
it is necessary to find a giant high-order Kerr nonlinearity that can be realized at very weak light level~\cite{Scully}.

In this article, we make a systematic theoretical investigation on the optical Kerr effect in a cold Rydberg atomic system via EIT. By using an approach beyond mean-field theory~\cite{Sevincli,stan,Sche} on the correlators of one-body, two-body, and three-body based on a second-order ladder approximation, we show that the system possesses not only an enhanced third-order nonlinear optical susceptibility, but also a giant fifth-order nonlinear optical susceptibility, which has a cubic dependence on atomic density and can be arrived at the
order of magnitude $10^{-11}$\,m$^4$V$^{-4}$. Our results demonstrate that both the third-order and the fifth-order nonlinear optical susceptibilities consist of two parts. One part is contributed by photon-atom interaction and another part comes from the Rydberg-Rydberg interaction. The Kerr nonlinearity induced by the Rydberg-Rydberg interaction plays a leading role at high atomic density. We find that the fifth-order  nonlinear optical susceptibility in the Rydberg-EIT system may be five orders of magnitude larger than that obtained in traditional EIT systems, which may have promising applications in light and quantum information processing and transmission at weak-light level.

Before preceding, we note that third-order Kerr nonlinearity was considered in \cite{Sevincli,stan,Sche} where nonlinearity is estimated by using the approach beyond mean-field theory, and in \cite{Sche} where a second-order ladder approximation is adopted to investigate coherent population trapping in Rydberg atoms. Furthermore, the interaction between Rydberg atoms via EIT was also suggested in \cite{Gorshkov,Petrosyan}. However, our work is different from \cite{Sevincli,stan,Sche,Gorshkov,Petrosyan}. First, no fifth-order Kerr nonlinearity was considered in \cite{Sevincli,stan,Sche,Gorshkov,Petrosyan} (see also recent review~\cite{Pri}). Second,
our study (see below) shows that both the photon-atom interaction and the Rydberg-Rydberg interaction have significant contributions to the Kerr nonlinearities (including third-order and fifth-order ones), but the contribution of the photon-atom interaction was overlooked in \cite{Sevincli,stan,Sche,Gorshkov,Petrosyan}.

The remainder of the article is arranged as follows. In Sec.~\ref{sec2}, the physical model of the Rydberg-EIT system under study is described.  In Sec.~\ref{sec3}, a perturbation expansion is used to solve the equations of motion of many-body correlators. In Sec.~\ref{sec4}, explicit expressions of the nonlinear optical susceptibilities are presented. Finally, the last section contains a summary of the main results of our work.

\section{Model}\label{sec2}

\begin{figure}
\centering
\includegraphics[width=0.9\textwidth]{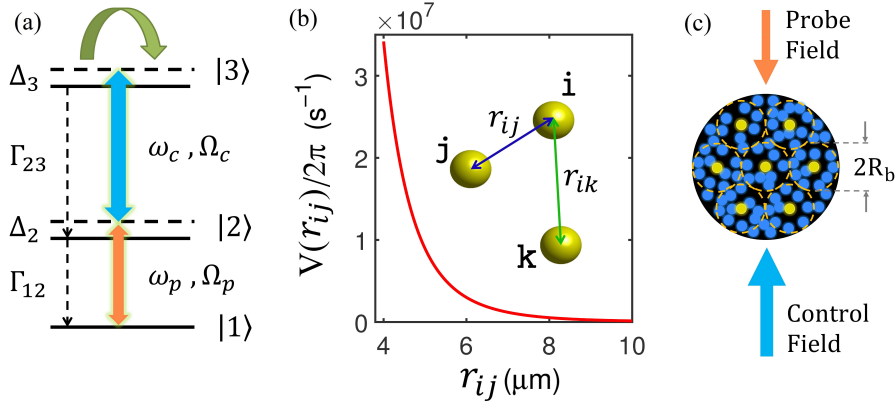}
\caption{\footnotesize (a) Excitation scheme of the three-level ladder system, in which the probe field with angular frequency $\omega_p$ and half Rabi frequency $\Omega_p$ couples the levels $|1\rangle$ and $|2\rangle$, and the control field with  angular frequency $\omega_c$ and half Rabi frequency $\Omega_c$ couples the levels $|2\rangle$ and
$|3\rangle$. $\Delta_2$ and $\Delta_3$ are one- and two-photon detunings, respectively; $\Gamma_{12}$ ($\Gamma_{23}$) is the spontaneous emission decay rate from $|2\rangle$ to $|1\rangle$ ($|3\rangle$ to $|2\rangle$). (b) The long-range interaction potential of $^{87}$Rb atoms $V(r_{ij})=-C_6/r_{ij}^6$ (red solid line) as a function of $r_{ij}=|{\bf r}_i-{\bf r}_j|$, describing the interaction between the atom at ${\bf r}_i$ and the atom at ${\bf r}_j$ (represented by yellow spheres), both of which are at the Rydberg state $|3\rangle=|60\,S_{1/2}\rangle$. (c) Schematic of Rydberg blockade. The long-range interaction between Rydberg atoms blocks the excitation of the atoms within  blockade spheres (i.e. the ones with the boundary indicated by the yellow dashed lines) of radius $R_b$. In each blocked sphere only one Rydberg atom (small yellow sphere) is excited and other atoms (small blue spheres) are prevented to be excited. The orange (blue) arrow indicates the propagating direction of the probe (control) field. }\label{fig1}
\end{figure}

We consider an ensemble of lifetime-broadened three-level atomic gas with a ladder-type level
configuration, shown schematically in Fig.~\ref{fig1}(a).
%
%
We assume the atomic gas are loaded into a magneto-optical trap and works at a ultracold temperature
so that their center-of-mass motion is negligible.  A weak probe field of angular frequency $\omega_p$ (half Rabi frequency $\Omega_p$) couples to the transition between $|1\rangle$ and $|2\rangle$, and a strong control field of angular
frequency $\omega_c$ (half Rabi frequency $\Omega_c$) couples to the transition
between $|2\rangle$ and $|3\rangle$. The electric field of the system can be written as
${\bf E}({\bf r}, t)={\bf E}_p({\bf r}, t)+{\bf E}_c({\bf r}, t)$
with ${\bf E}_p({\bf r}, t)={\bf e}_p\, \mathcal{E}_p \,\exp [i({\bf k}_p
\cdot {\bf r}-\omega_p t)]+{\rm c.c.}$
and ${\bf E}_c({\bf r}, t)={\bf e}_c\, \mathcal{E}_c \,\exp [i({\bf k}_c \cdot {\bf r}-\omega_c t)]+{\rm c.c.}$, where c.c. represents
complex conjugate and ${\bf k}_p$, ${\bf e}_p$  and $\mathcal{E}_p$  ( ${\bf k}_c$, ${\bf e}_c$ and $\mathcal{E}_c$)
are respectively the wavevector, polarization unit vector and amplitude of the probe field
(control) field. The upper state $|3\rangle$ is chosen as a Rydberg state, which is assumed to be  $|3\rangle=|60\,S_{1/2}\rangle$
for simplicity. The atoms in Rydberg states (i.e. Rydberg atoms)
have many exaggerated properties, including long radiative lifetime, large electric dipole moment, strong Rydberg-Rydberg interaction, and so on~\cite{Gal}.

Under electric-dipole and rotating-wave approximations, in the Heisenberg picture the Hamiltonian of the atomic gas including
the Rydberg-Rydberg interaction is given by $\hat{H}_H={\cal N}_{a}\int {\rm d}^3 \mathbf{r} \,\hat{\cal H}_H(\mathbf{r},t)$, with
${\cal N}_{a}$ the atomic density and $\hat{\cal H}_H(\mathbf{r},t)$ the Hamiltonian of the atom at position ${\bf r}$ of the form
\begin{eqnarray} \label{eq1}
\hat{{\cal H}}_H(\mathbf{r},t) &=
&  \sum_{\alpha=1}^{3}\hbar\omega_{\alpha}\hat{S}_{\alpha\alpha}(\mathbf{r},t)
-\hbar\left[\Omega_p\hat{S}_{12}(\mathbf{r},t)
+\Omega_p^\ast\hat{S}_{21}(\mathbf{r},t)+\Omega_c\hat{S}_{23}(\mathbf{r},t)
+\Omega_c^\ast\hat{S}_{32}(\mathbf{r},t)\right]
\nonumber\\
& & +{\cal N}_{a}\int{\rm d}^3{\mathbf{r}^\prime}\hat{S}_{33}(\mathbf{r}^\prime,t)\hbar V(\mathbf{r}^\prime-\mathbf{r})\hat{S}_{33}
(\mathbf{r},t),
\end{eqnarray}
where $\hbar \omega_{\alpha}$ the eignenergy of the state $|\alpha\rangle$, $\Omega_p=(\mathbf{e}_p\cdot \mathbf{p}_{21})\mathcal{E}_p/\hbar$ and $\Omega_c=(\mathbf{e}_c\cdot \mathbf{p}_{32})\mathcal{E}_c/\hbar$ are respectively the half Rabi frequencies of the probe and control fields with $\mathbf{p}_{\alpha\beta}$ the electric dipole matrix element associated with the
transition from $|\beta\rangle$ to $|\alpha\rangle$,
$\hat{S}_{\alpha\beta}=|\beta\rangle\langle \alpha|e^{i[(\mathbf{k}_{\beta}-\mathbf{k}_{\alpha})\cdot\mathbf{r}-(\omega_{\beta}
-\omega_{\alpha}+\Delta_{\beta}-\Delta_{\alpha})t]}$
are transition operators $(\alpha,\beta = 1,2,3)$  satisfying the commutation relation
\begin{equation}
\left[\hat{S}_{\alpha\beta}(\mathbf{r},t),\hat{S}_{\mu\nu}(\mathbf{r}^{\prime},t)\right]
=\left(\delta_{\alpha\nu}\hat{S}_{\mu \beta}(\mathbf{r},t)-\delta_{\mu \beta}\hat{S}_{\alpha\nu}(\mathbf{r}^{\prime},t)\right)\delta_{\mathbf{r} \mathbf{r}^{\prime}},
\end{equation}
where $\delta_{\alpha\beta}$ is Kronecker symbol. The last term on the right side of  Eq.~(\ref{eq1}) is the contribution of the Rydberg-Rydberg interaction, i.e. the Rydberg atom at position $\mathbf{r}$ interacts with the Rydberg atom at position $\mathbf{r}^\prime$ described by the long-range interaction potential $V(\mathbf{r}^\prime-\mathbf{r})$. Because the Rydberg-Rydberg interaction results in a Rydberg blockade~\cite{Saf,Pri1,Sevincli0}, the integration region of the radial coordinate $r^{\prime}$ in the integral in the last line of Eq.~(\ref{eq1}) is from $2R_b$ to infinity, where $R_b$ is the radius of Rydberg blockade sphere [see Fig.~\ref{fig1}(c)].

The equations of motion for one-body density matrix $\rho$ is given by~\cite{Sevincli0,stan}
\begin{subequations} \label{eq2}
\begin{eqnarray}
&&i\frac{\partial }{\partial t}\rho_{11}-i\Gamma_{12}\rho
_{22}-\Omega
_{p}\rho_{12}+\Omega _{p}^{\ast}\rho_{21}=0,\label{eq21}\\
&&i\frac{\partial }{\partial t}\rho_{22}-i\Gamma_{23}\rho
_{33}+i\Gamma_{12}\rho _{22}+\Omega _{p}\rho_{12}-\Omega
_{p}^{\ast}\rho_{21}-\Omega_{c}\rho_{23}+\Omega_{c}^{\ast
}\rho_{32}=0,\label{eq22}\\
&&i\frac{\partial }{\partial t}\rho_{33}+i\Gamma_{23}\rho
_{33}+\Omega_{c}\rho_{23}-\Omega_{c}^{\ast }\rho_{32}=0,\label{eq23}\\
&&\left(i\frac{\partial }{\partial t}+d_{21}\right)
\rho_{21}-\Omega_{p}(\rho_{22}-\rho_{11})+\Omega_{c}^{\ast
}\rho_{31}=0,\label{eq24}\\
&&\left(i\frac{\partial }{\partial t}+d_{31}\right) \rho_{31}-\Omega
_{p}\rho_{32}+\Omega_{c}\rho_{21}-{\cal
N}_a\int{d^3\mathbf{r}^\prime V(\mathbf{r}^\prime-\mathbf{r})\rho\rho_{33,31}(\mathbf{r}^\prime,\mathbf{r},t)}=0,\label{eq25}\\
&&\left(i\frac{\partial }{\partial t}+d_{32}\right) \rho_{32}-\Omega_{p}^{\ast}\rho_{31}-\Omega_{c}(\rho_{33}-\rho_{22})\nonumber\\
&& \hspace{0.8cm}-{\cal
N}_a\int{d^3\mathbf{r}^\prime V(\mathbf{r}^\prime-\mathbf{r})\rho\rho_{33,32}(\mathbf{r}^\prime,\mathbf{r},t)}=0,\label{eq26}
\end{eqnarray}
\end{subequations}
where $\rho_{\alpha\beta}=\langle\hat{S}_{\alpha\beta}\rangle$~\cite{note01} is the one-body density matrix element,
$d_{\alpha\beta}=\Delta_{\alpha}-\Delta_{\beta}+i\gamma_{\alpha\beta}$ ($\Delta_1=0$;
$\alpha, \beta= 1, 2, 3;\alpha\neq \beta)$,
$\Delta_2=\omega_p-(\omega_2-\omega_1)$ and $\Delta_3=\omega_p+\omega_c-(\omega_3-\omega_1)$ are respectively the one-photon and two-photon detunings,
$\gamma_{\alpha\beta}=(\Gamma_\alpha+\Gamma_\beta)/2+\gamma_{\alpha\beta}^{\rm col}$  with $\Gamma_\beta=\sum_{\alpha<\beta}\Gamma_{\alpha\beta}$. Here $\Gamma_{\alpha\beta}$ denotes the spontaneous emission decay rate from the state $|\beta\rangle$ to the state $|\alpha\rangle$
and $\gamma_{\alpha\beta}^{\rm col}$  represents the dephasing rate reflecting the loss of phase coherence between $|\alpha\rangle$ and $|\beta\rangle$.

From the above equations we see that there are
two evident nonlinear characters in the system: (i)~There is a photon-atom interaction due to the resonant coupling between
the probe field and the atoms even when the Rydberg-Rydberg interaction is absent. (ii)~There is an atom-atom interaction reflected by
the last terms on the left hand side of Eq.~\ref{eq2}(e) and Eq.~\ref{eq2}(f), i.e. the two-body density matrix elements (or the two-body correlators) $\rho\rho_{33,3\alpha}(\mathbf{r}^\prime,\mathbf{r},t)\equiv\langle \hat{S}_{33}
(\mathbf{r}^\prime,t)\hat{S}_{3\alpha}(\mathbf{r},t)\rangle$ ($\alpha=1,2$) contributed from the Rydberg-Rydberg interaction. It is just these two different nonlinear characters that make the Rydberg-EIT system possess very interesting nonlinear optical properties. Especially, two different types of Kerr nonlinearities (one is resulted from the photon-atom interaction and another one is resulted from the Rydberg-Rydberg interaction) occur in the system, as will be illustrated below.

Our model can be easily realized by experiment. One of candidates is the laser-cooled $^{87}$Rb atomic gas with the atomic states
shown in Fig.~{\ref{fig1}(a) assigned as~\cite{data,pritchard2}
\begin{eqnarray}\label{param}
& & \label{param1}|1\rangle=|5s^2 S_{1/2},F=2\rangle,\,\,\,\,|2\rangle=|5p^2 P_{3/2},F=3\rangle,
\nonumber \,\,\,\,\\
& &  \label{param2}|3\rangle=|ns^2 S_{1/2}\rangle,\,\,\,\,\Gamma_{12}=2\pi\times6\,\,{\rm MHz}, \,\,\,\,\Gamma_{23}=2\pi\times3 \nonumber \,\, {\rm kHz},
\end{eqnarray}
with $n$ principle quantum number and other parameters taken as $\Omega_c=2\pi\times32\,\,${\rm MHz},  $\Delta_2=2\pi\times160\,\,${\rm MHz}. All calculations given below will be based on these realistic physical parameters.
The long-range interaction potential between two Rydberg atoms has the form as $V(r_{ij})=-C_6/r_{ij}^6$\cite{pritchard2}, where $r_{ij}=|{\bf r}_i-{\bf r}_j|$ is the distance between the $i$th and $j$th Rydberg atoms, represented by the yellow spheres in Fig.~\ref{fig1}(b).  The red solid line in Fig.~\ref{fig1}(b) is the curve of the
long-range interaction potential $-C_6/r_{ij}^6$  as a function of $r_{ij}$ for $n=60$, with the dispersion parameter $C_6\simeq-2\pi\times140\,{\rm GHz\,\mu m}^6$, adopted from \cite{pritchard2}.

Due to the Rydberg-Rydberg interaction, an atom in the state $|3\rangle$ would induce an energy-shift $V(R)$ of the state $|3\rangle$ of another atom separated by distance $R$, which translates into an effective two-photon detuning. Then the long-range interaction energy-shift will block the excitation of all the atoms for which $V(R)\geq\delta_{\rm EIT}$, where $\delta_{\rm EIT}$ is
the linewidth of EIT transmission spectrum (i.e. the width of EIT transparency window), defined by $\Omega_c^2/\gamma_{12}$ for $\Delta_2=0$ and $\Omega_c^2/|\Delta_2|$ for $|\Delta_2|\gg\gamma_{12}$\,
(we assumed $-\Delta_2/C_6>0$). Thus the blockade sphere has the radius $R_b=(|C_6/\delta_{\rm EIT}|)^{1/6}\simeq5.29{\rm \mu m}$\,\cite{pritchard2,Gorshkov,note000}. Comparing this to the average interatomic separation obtained by $\bar{R}=(5/9){\cal N}_a^{-1/3}\simeq2.5{\rm \mu m}$ for ${\cal N}_a=10^{10}$cm${^{-3}}$, the blockade effect can be obviously observed, as shown in Fig.~\ref{fig1}(c). The system can be divided into many blockade spheres (represented by the spheres with the boundary indicated by yellow dashed line in Fig.~\ref{fig1}(c)\,) and each blockade sphere contains only one Rydberg atom (represented by the small yellow sphere in Fig.~\ref{fig1}(c)). Hence, the spatial coarse-graining distance between two nearest Rydberg atoms has size $2R_{b}$\cite{Petrosyan}.

We are interested in the optical Kerr effects, especially the third-order and fifth-order nonlinear optical susceptibilities of the system. To this aim, we need the relation between the optical susceptibility of the probe field and the density matrix elements. Since the total electric polarization intensity of the system is given by
${\mathbf P} = {\cal N}_a \sum_{\alpha,\beta=1}^{3} {\mathbf p}_{\alpha\beta} \rho_{\beta\alpha}
\exp \{i[({\bf k}_{\beta}-{\bf k}_{\beta})\cdot {\bf r}-( \omega_{\beta}-\omega_{\alpha}+\Delta_{\beta}-\Delta_{\alpha})t]\}$, the
electric polarization intensity of the probe field reads
${\mathbf P}_p={\cal N}_a\{ {\mathbf p}_{12}\rho_{21}{\rm exp}[i({\bf k}_p\cdot {\bf r}-\omega_p t)]+{\rm c.c.}\}$, by which one can
obtain the optical susceptibility $\chi_p$ of the probe field
by using the formula ${\mathbf P}_p=\varepsilon_0  \chi_p {\bf e}_p \mathcal{E}_p{\rm exp}[i({\bf k}_p\cdot {\bf r}-\omega_p t)]+{\rm c.c.}$,  which yields
\begin{equation}\label{susceptibilty0}
\chi_p=\frac{ {\cal N}_a ( {\bf e}_p\cdot {\bf p}_{12}) \rho_{21} }{ \varepsilon_0 \mathcal{E}_{p} }.
\end{equation}

To obtain the explicit expression of $\rho_{21}$, we must solve Eqs. (\ref{eq21})-(\ref{eq26}).
However, due to the Rydberg-Rydberg interaction,
we must also solve the motion of equations for the two-body correlators $\langle \hat{S}_{33}\hat{S}_{31}\rangle$ and
$\langle\hat{S}_{33}\hat{S}_{32}\rangle$ simultaneously
\begin{subequations} \label{correlators}
\begin{eqnarray}
&&\label{correlators1} \left(i\frac{\partial }{\partial t}+d_{31}+i\Gamma_{23}-V(\mathbf{r}^{\prime}-\mathbf{r})\right)\langle\hat{S}_{33}\hat{S}_{31}\rangle
+{\Omega}
_{c}\left(\langle\hat{S}_{23}\hat{S}_{31}\rangle+\langle\hat{S}_{33}\hat{S}_{21}\rangle\right)
-\Omega_c^\ast \langle\hat{S}_{32}\hat{S}_{31}\rangle \nonumber\\
&&\hspace{0.5cm} -\Omega_p \langle\hat{S}_{33}\hat{S}_{32}\rangle-{\cal N}_a\int{d^3\mathbf{r}^{\prime\prime}\langle\hat{S}_{33}(\mathbf{r}^{\prime\prime},t)\hat{S}_{33}
(\mathbf{r}^{\prime},t)\hat{S}_{31}(\mathbf{r},t)\rangle V(\mathbf{r}^{\prime\prime}-\mathbf{r})}=0,\\
&&\label{correlators2} \left(i\frac{\partial }{\partial t}+i\Gamma_{23}+d_{32}-V(\mathbf{r}^{\prime}-\mathbf{r})\right)\langle\hat{S}_{33}\hat{S}_{32}\rangle
+\Omega_{c}\left(\langle\hat{S}_{23}\hat{S}_{32}\rangle-\langle\hat{S}_{33}\hat{S}_{33}\rangle
+\langle\hat{S}_{33}\hat{S}_{22}\rangle\right)-\hat{\Omega}_{c}^\ast \langle\hat{S}_{32}\hat{S}_{32}\rangle\nonumber\\
&&\hspace{0.5cm} -\Omega_{p}^* \langle\hat{S}_{33}\hat{S}_{31}\rangle-{\cal N}_a\int{d^3\mathbf{r}^{\prime\prime}\langle\hat{S}_{33}(\mathbf{r}^{\prime\prime},t)
\hat{S}_{33}(\mathbf{r}^{\prime},t)\hat{S}_{32}(\mathbf{r},t)\rangle V(\mathbf{r}^{\prime\prime}-\mathbf{r})}=0,
\end{eqnarray}
\end{subequations}
where $\mathbf{r}^{\prime}\neq\mathbf{r}^{\prime\prime}$ and $\hat{S}_{\alpha\beta}\hat{S}_{\mu\nu}$
in the terms without integration means $\hat{S}_{\alpha\beta}(\mathbf{r}^{\prime},t)\hat{S}_{\mu\nu}(\mathbf{r},t)$.

Eqs. (\ref{correlators1}) and
(\ref{correlators2}) have the following features. (i)~The equations for two-body correlators $\langle\hat{S}_{33}\hat{S}_{3\alpha}\rangle$ $(\alpha=1,2)$
involve many other two-body correlators (e.g. $\langle\hat{S}_{23}\hat{S}_{31}\rangle$, etc.). Thus one also have to solve additional equations of other two-body correlators. An explicit  list of the equations of motion for two-body correlators $\langle\hat{S}_{\alpha\beta}\hat{S}_{\mu\nu}\rangle$ of the system is too long and omitted here.  (ii)~The equations for the two-body correlators involve three-body correlators (e.g. $\langle\hat{S}_{33}\hat{S}_{33}\hat{S}_{31}\rangle$, etc.),
which obey the equations of motion for three-body correlators (which are lengthy and not listed here)
and also have to be solved. Similarly, the equations of motion of the three-body correlators involve four-body correlators.
Finally, one obtains an infinite hierarchy of equations of motion for the correlators of one-body, two-bodies, three-bodies, and so on. Obviously, to make the problem tractable one must truncate the hierarchy of the equations for many-body correlators by using an appropriate method. Here we adopt a second-order ladder approximation, such that for moderate atomic density the three-body correlation terms in the two-body correlator equations are factorized in the following way~\cite{Sevincli,stan,Sche,Mukamel}
\begin{eqnarray}\label{factorize}
 & & \langle S_{\alpha\beta}(\mathbf{r}^{\prime\prime})S_{\mu\nu}(\mathbf{r}^{\prime})
 S_{\alpha^\prime\beta^\prime}(\mathbf{r}) \rangle \nonumber\\
 & & \hspace{0.0cm} =\langle S_{\alpha\beta}(\mathbf{r}^{\prime\prime})\rangle\langle S_{\mu\nu}(\mathbf{r}^{\prime})S_{\alpha^\prime\beta^\prime}(\mathbf{r}) \rangle+\langle S_{\alpha\beta}(\mathbf{r}^{\prime\prime})S_{\mu\nu}(\mathbf{r}^{\prime})\rangle\langle S_{\alpha^\prime\beta^\prime}(\mathbf{r}) \rangle, \nonumber\\
 & &\hspace{0.3cm}+\langle S_{\alpha\beta}(\mathbf{r}^{\prime\prime})S_{\alpha^\prime\beta^\prime}(\mathbf{r})\rangle\langle S_{\mu\nu}(\mathbf{r}^{\prime}) \rangle-2\langle S_{\alpha\beta}(\mathbf{r}^{\prime\prime})\rangle \langle S_{\mu\nu}(\mathbf{r}^{\prime})\rangle \langle S_{\alpha^\prime\beta^\prime}(\mathbf{r}) \rangle,
\end{eqnarray}
As a special case, when the atomic density is low and the interaction between atoms is weak so that the correlation between atoms are negligible, one has
$\langle S_{\alpha\beta}(\mathbf{r}^{\prime\prime})S_{\mu\nu}(\mathbf{r}^{\prime})S_{\alpha^\prime\beta^\prime}(\mathbf{r}) \rangle\rightarrow\langle S_{\alpha\beta}(\mathbf{r}^{\prime\prime})\rangle \langle S_{\mu\nu}(\mathbf{r}^{\prime})\rangle \langle S_{\alpha^\prime\beta^\prime}(\mathbf{r}) \rangle$, corresponding to a mean-field approximation. We stress that
the mean-field approximation is not valid for the Rydberg gases even at lower atomic density because of the strong Rydberg-Rydberg interaction. On the other hand, to acquire a giant nonlinear optical effect, a higher atomic density is usually needed and hence one must adopt a method beyond the mean-field approximation~\cite{note111}.
The factorization method stated above is an effective approach for dealing with interacting multi-body problems and has been widely adopted in nonlinear laser spectroscopy~\cite{Mukamel} and Bose-condensed gases~\cite{Griffin}, by which the equations of motion for one-body and two-body correlators are closed and hence can be solved by using some suitable techniques.

\section{Solutions based on perturbation expansion}\label{sec3}

Although by using the factorization method stated above the equations of motion for one-body
and two-body correlations can be made to be closed and their number becomes finite, they are still
nonlinear due to the coupling with the applied laser field. Fortunately, since the probe field in
the EIT-based experiments~\cite{Moha,Mohapatra,pritchard1,pritchard2,Parigi,Hofmann,Maxwell2013} is weak we hence can make a
perturbation expansion of the correlators in the powers of the Rabi frequency of probe field $\Omega_p$~\cite{stan} to solve these nonlinear equations in a systematic way. In fact, when EIT systems are weakly driven, the half Rabi frequency $\Omega_p$ is a natural expansion parameter for investigating many weak nonlinear phenomena, including ultraslow and weak-light solitons in EIT-based systems~\cite{Huang,Hang1,Chen}.

To investigate the optical Kerr effects in the present Rydberg-EIT system, we make the perturbation expansion~~\cite{stan}
$\rho_{\alpha1}=\Omega_p\sum_{l=0}\rho_{\alpha1}^{(2l+1)}|\Omega_p|^{2l}$, $\rho_{32}=\sum_{l=1}\rho_{32}^{(2l)}|\Omega_p|^{2l}$, $\rho_{\beta\beta}=\sum_{l=0}\rho_{\beta\beta}^{(2l)}|\Omega_p|^{2l}$
with $\rho_{\beta\beta}^{(0)} =\delta_{\beta1}\delta_{\beta1}$($\alpha=2,3; \beta=1,2,3$). Substituting this expansion into the Eq.~(\ref{eq2}) for the one-body density matrix elements
and comparing the expansion parameter of each power $\Omega_p$, we obtain a set of approximated equations for
$\rho_{\alpha\beta}^{(l)}$ , which are listed in Appendix~\ref{ap2}. The approximated equations for the two-body density matrix (correlator) elements $\rho\rho_{\alpha\beta,\mu\nu}^{(l)}$ after using the factorization formula~(\ref{factorize})
can also be obtained, which are listed in Appendix~\ref{ap3}. In order to acquire third-order and fifth-order
nonlinear optical susceptibilities, we must solve the expansion equations from the first order to the fifth order.
Although these expansion equations are lengthy and complicated, they become linear after the above expansion and thus can be solved analytically order by order in a systematical and clear way. Notice that in this work we are interested in static (or instantaneous)
nonlinear optical susceptibilities, both the probe and control fields are assumed to be continuous waves. Thus the operator $\partial/\partial t$ in all equations of the correlators can be put into zero.

At the first order ($l=1$), we obtain the solution $\rho_{21}^{(1)}=d_{31}/D$ and $\rho_{31}^{(1)}=-\Omega_c/D$,
with $D=|\Omega_c|^2-d_{21}d_{31}$. For the second order ($l=2$), one obtains the solution
\begin{subequations} \label{secondorder}
\begin{eqnarray}
&& \rho_{11}^{(2)}=\frac{[i\Gamma_{23}-2|\Omega_c|^2 M]N -i\Gamma_{12}\left(\frac{|\Omega_c|^2}{D^{\ast}d_{32}^{\ast}}
-\frac{|\Omega_c|^2}{Dd_{32}}\right)}{-\Gamma_{12}\Gamma_{23}-i\Gamma_{12}|\Omega_c|^2 M},\\
&& \rho_{33}^{(2)}=\frac{1}{i\Gamma_{12}}\left(N-i\Gamma_{12}\rho_{11}^{(2)}\right),\\
&& \rho_{32}^{(2)}=\frac{1}{d_{32}}\left(-\frac{\Omega_c}{D}+2\Omega_c\rho_{33}^{(2)}+\Omega_c\rho_{11}^{(2)}\right),
\end{eqnarray}
\end{subequations}
where $M=1/d_{32}-1/d_{32}^{\ast}$, $N=d_{31}^{\ast}/D^{\ast}-d_{31}/D$. At the third order ($l=3$), the solution reads
$\rho_{21}^{(3)}=a_{21}^{(3)}+{\cal N}_ab_{21}^{(3)}$ and $\rho_{31}^{(3)}=a_{31}^{(3)}+{\cal N}_ab_{31}^{(3)}$, with
\begin{subequations} \label{thirdorder}
\begin{eqnarray}
&& a_{21}^{(3)}=\frac{\Omega_c^{\ast}\rho_{32}^{(2)}+d_{31}(2\rho_{11}^{(2)}+\rho_{33}^{(2)})}{|\Omega_c|^2-d_{21}d_{31}},\\
&& b_{21}^{(3)}=\frac{\Omega_c^\ast\int{d^3\mathbf{r}^\prime \rho\rho_{33,31}^{(3)}(\mathbf{r}^\prime-\mathbf{r})V(\mathbf{r}^\prime-\mathbf{r})}}{|\Omega_c|^2-d_{21}d_{31}},\\
&& a_{31}^{(3)}=\frac{-(2\rho_{11}^{(2)}+\rho_{33}^{(2)})\Omega_c+d_{21}\rho_{32}^{(2)}}{|\Omega_c|^2-d_{21}d_{31}},\\
&& b_{31}^{(3)}=-\frac{\int{d^3\mathbf{r}^\prime \rho\rho_{33,31}^{(3)}(\mathbf{r}^\prime-\mathbf{r})V(\mathbf{r}^\prime-\mathbf{r})}d_{21}}{|\Omega_c|^2-d_{21}d_{31}},
\end{eqnarray}
\end{subequations}
where a general expression of $\rho\rho_{33,31}^{(3)}$  is given in Appendix~\ref{ap3} [see Eq.~(\ref{order33313})\,].
Because we have assumed the probe field to be weak, the Rydberg-Rydberg interaction gives the contribution to the
solution starting only from the third order approximation.

At the fourth order ($l=4$), the solution is given by $\rho_{11}^{(4)}=a_{11}^{(4)}+{\cal N}_ab_{11}^{(4)}$,
$\rho_{33}^{(4)}=a_{33}^{(4)}+{\cal N}_ab_{33}^{(4)}$, and $\rho_{32}^{(4)}=a_{32}^{(4)}+{\cal N}_ab_{32}^{(4)}$,
with
\begin{subequations} \label{fourthorder}
\begin{eqnarray}
&& a_{11}^{(4)}=\frac{[i\Gamma_{23}-2|\Omega_c|^2 M](a_{21}^{\ast(3)}-a_{21}^{(3)}) +i\Omega_cd_{32}^{\ast-1}\Gamma_{12}a_{31}^{\ast(3)}-i\Omega_c^\ast d_{32}^{-1}\Gamma_{12}a_{31}^{(3)}}{-\Gamma_{12}\Gamma_{23}-
i\Gamma_{12}|\Omega_c|^2 M},\\
&& b_{11}^{(4)}=\frac{[i\Gamma_{23}-2|\Omega_c|^2 M](b_{21}^{\ast(3)}-b_{21}^{(3)}) +i\Omega_cd_{32}^{\ast-1}\Gamma_{12}b_{31}^{\ast(3)}-i\Omega_c^\ast d_{32}^{-1}\Gamma_{12}b_{31}^{(3)}}{-\Gamma_{12}\Gamma_{23}-
i\Gamma_{12}|\Omega_c|^2 M}\\
&&\hspace{1cm}+\int{d^3\mathbf{r}^\prime \frac{i\Gamma_{12}\Omega_cd_{32}^{\ast-1}\rho\rho_{33,32}^{\ast(4)}
(\mathbf{r}^\prime-\mathbf{r})V(\mathbf{r}^\prime-\mathbf{r})-i\Gamma_{12}\Omega_c^\ast d_{32}^{-1}\rho\rho_{33,32}^{(4)}(\mathbf{r}^\prime-\mathbf{r})V(\mathbf{r}^\prime-\mathbf{r})}{-\Gamma_{12}\Gamma_{23}-
i\Gamma_{12}|\Omega_c|^2 M}}, \nonumber\\
&& a_{33}^{(4)}=-ia_{21}^{\ast(3)}/\Gamma_{12}-a_{11}^{(4)}+ia_{21}^{(3)}/\Gamma_{12},\\
&& b_{33}^{(4)}=-ib_{21}^{\ast(3)}/\Gamma_{12}-b_{11}^{(4)}+ib_{21}^{(3)}/\Gamma_{12},\\
&&a_{32}^{(4)}=\frac{1}{d_{32}}\left(a_{31}^{(3)}
+2\Omega_ca_{33}^{(4)}+\Omega_ca_{11}^{(4)}\right),\\
&&b_{32}^{(4)}=\frac{1}{d_{32}}\left(b_{31}^{(3)}
+2\Omega_cb_{33}^{(4)}+\Omega_cb_{11}^{(4)}\right)+\int{d^3\mathbf{r}^\prime
\frac{\rho\rho_{33,32}^{(4)}(\mathbf{r}^\prime-\mathbf{r})V(\mathbf{r}^\prime-\mathbf{r})}{d_{32}}}.
\end{eqnarray}
\end{subequations}

With the above solutions, we go to the fifth order ($l=5$). The solution at this order reads
$\rho_{21}^{(5)}=a_{21}^{(5)}+{\cal N}_ab_{21}^{(5)}+{\cal N}_a^2c_{21}^{(5)}$ and $\rho\rho_{33,31}^{(5)}=aa_{33,31}^{(5)}+{\cal N}_abb_{33,31}^{(5)}$
with
\begin{subequations} \label{fifthorder}
\begin{eqnarray}
&& a_{21}^{(5)}=\frac{\Omega_c^{\ast}a_{32}^{(4)}+d_{31}(2a_{11}^{(4)}+a_{33}^{(4)})}{|\Omega_c|^2-d_{21}d_{31}},\\
&& b_{21}^{(5)}=\frac{\Omega_c^{\ast}b_{32}^{(4)}+d_{31}(2b_{11}^{(4)}+b_{33}^{(4)})}{|\Omega_c|^2-d_{21}d_{31}}
+\frac{\Omega_c^\ast\int{d^3\mathbf{r}^\prime  aa_{33,31}^{(5)}(\mathbf{r}^\prime-\mathbf{r})V(\mathbf{r}^\prime-\mathbf{r})}}{|\Omega_c|^2-d_{21}d_{31}},\\
&& c_{31}^{(5)}=\frac{\Omega_c^\ast\int{d^3\mathbf{r}^\prime bb_{33,31}^{(5)}(\mathbf{r}^\prime-\mathbf{r})V(\mathbf{r}^\prime-\mathbf{r})}}{|\Omega_c|^2-d_{21}d_{31}},
\end{eqnarray}
\end{subequations}
where the expressions of $\rho\rho_{33,32}^{(4)}$, $aa_{33,31}^{(5)}$ and $bb_{33,31}^{(5)}$
in Eqs.~(\ref{fourthorder}) and (\ref{fifthorder}) have been given in Appendix \ref{ap3}
[see Eqs.~(\ref{aa33324}), (\ref{aa33315}), and (\ref{bb33315})\,].

\section{Giant third-order and fifth-order nonlinear optical susceptibilities}\label{sec4}

Collecting the first-order to the fifth-order solutions of $\rho_{21}$ obtained in the last section,  we obtain $\rho_{21}\simeq \rho_{21}^{(1)}\Omega_p+\rho_{21}^{(3)}|\Omega_p|^2 \Omega_p+\rho_{21}^{(5)}|\Omega_p|^5 \Omega_p+...$,
where $\rho_{21}^{(j)}$ ($j=1,3,5,...$) are independent of $\Omega_p$ and their explicit expressions have been given
in the previous section.  Using the formula (\ref{susceptibilty0}) and the
definition $\Omega_p=({\bf e}_p\cdot {\bf p}_{21}){\mathcal E}_p/\hbar$, we have
\begin{equation}\label{susceptibilty}
\chi_p\simeq \chi_p^{(1)}+\chi_p^{(3)}|\mathcal{E}_p|^2+\chi_p^{(5)}|\mathcal{E}_p|^4,
\end{equation}
where $\chi_p^{(1)}$, $\chi_p^{(3)}$, and $\chi_p^{(5)}$ are respectively the first-order (linear), the third-order and the
fifth-order (nonlinear) optical susceptibilities of the probe field, defined by
\begin{subequations}\label{chi30}
\begin{eqnarray}
& & \chi_{p}^{(1)}=\frac{{\cal N}_a|\mathbf{p}_{12}|^2}{\varepsilon_0\hbar }\frac{d_{31}}{D},\\
& & \chi_p^{(3)}=\chi_{p1}^{(3)}+\chi_{p2}^{(3)},\\
& & \chi_p^{(5)}=\chi_{p1}^{(5)}+\chi_{p2}^{(5)},
\end{eqnarray}
\end{subequations}
with
\begin{subequations}\label{chi3}
\begin{eqnarray}
\chi_{p1}^{(3)}=&&\frac{{\cal N}_a|\mathbf{p}_{12}|^4}{\varepsilon_0\hbar^3 }\frac{1}{D}\left[\Omega_c^{\ast}\rho_{32}^{(2)}+d_{31}(2\rho_{11}^{(2)}+\rho_{33}^{(2)})\right],
\label{chip13}\\
\chi_{p2}^{(3)}=&&\frac{{\cal N}_a^2|\mathbf{p}_{12}|^4}{\varepsilon_0\hbar^3 }\frac{\Omega_c^\ast}{D}\int{d^3\mathbf{r}^\prime \rho\rho_{33,31}^{(3)}(\mathbf{r}^\prime-\mathbf{r})V(\mathbf{r}^\prime-\mathbf{r})},\label{chip23}\\
\chi_{p1}^{(5)}=&&\frac{{\cal N}_a|\mathbf{p}_{12}|^6}{\varepsilon_0\hbar^5 }\frac{1}{D}\left[\Omega_c^{\ast}a_{32}^{(4)}+d_{31}(2a_{11}^{(4)}+a_{33}^{(4)})\right],
\label{chip15}\\
\chi_{p2}^{(5)}=&&\frac{{\cal N}_a^2|\mathbf{p}_{12}|^6}{\varepsilon_0\hbar^5 }\frac{\Omega_c^{\ast}b_{32}^{(4)}+d_{31}(2b_{11}^{(4)}+b_{33}^{(4)})}{D}\nonumber\\
&&+\frac{{\cal N}_a^2|\mathbf{p}_{12}|^6}{\varepsilon_0\hbar^5 }\frac{\Omega_c^\ast}{D}\int{d^3\mathbf{r}^\prime aa_{33,31}^{(5)}(\mathbf{r}^\prime-\mathbf{r})V(\mathbf{r}^\prime-\mathbf{r})}\nonumber\\
&&+\frac{{\cal N}_a^3|\mathbf{p}_{12}|^6}{\varepsilon_0\hbar^5 }\frac{\Omega_c^\ast}{D}\int{d^3\mathbf{r}^\prime bb_{33,31}^{(5)}(\mathbf{r}^\prime-\mathbf{r})V(\mathbf{r}^\prime-\mathbf{r})},\label{chip25}
\end{eqnarray}
\end{subequations}
where $\chi_{p1}^{(3)}$ and $\chi_{p1}^{(5)}$ are the third-order and the fifth-order nonlinear optical susceptibilities arising from the interaction between the probe field and the atoms (i.e. by the photon-atom interaction), $\chi_{p2}^{(3)}$ and $\chi_{p2}^{(5)}$ are third-order and fifth-order nonlinear optical susceptibilities arising from the Rydberg-Rydberg interaction (i.e. by the atom-atom interaction). From the expressions (\ref{chip13}) and (\ref{chip15}), we see that
$\chi_{p1}^{(3)}$ and $\chi_{p1}^{(5)}$ have a linear dependence on the atomic density ${\cal N}_a$.
Differently, from the expressions (\ref{chip23}) and (\ref{chip25}) we observe that
$\chi_{p2}^{(3)}$ has a quadratic dependence on the atomic density (i.e. on ${\cal N}_a^2$) and
$\chi_{p2}^{(5)}$ has not only quadratic but also cubic dependence on the atomic density
(i.e. on ${\cal N}_a^2$ and ${\cal N}_a^3$), which implies that the nonlinear optical susceptibilities in the
Rydberg-EIT system are very sensitive to the change of the atomic density ${\cal N}_a$.

\begin{table}[!hbp]\label{table}
\caption{Real part ${\rm Re}(\chi_{p\alpha}^{(j)})$
and imaginary part ${\rm Im}(\chi_{p\alpha}^{(j)})$ ($j=3,5$; $\alpha=1,2$)
of the third-order and the fifth-order optical susceptibilities of
the Rydberg-EIT system obtained for the realistic system parameters given in the text.}
\centering
\begin{tabular}{c c c c}
\hline
\, & Real part & Imaginary part & Contributed by\\
\hline
$\chi_{p1}^{(3)}$ ~~~~~~& $-4.4\times10^{-11}$ m$^2$V$^{-2}$~~~~~~& $-1.3\times10^{-13}$ m$^2$V$^{-2}$~~~~~~& ~~~photon-atom interaction\\
$\chi_{p2}^{(3)}$ ~~~~~~& $-2.1\times10^{-8}$ m$^2$V$^{-2}$~~~~~~ & $-1.14\times10^{-10}$ m$^2$V$^{-2}$~~~~~~ & Rydberg-Rydberg interaction \\
$\chi_{p1}^{(5)}$ ~~~~~~& $2.6\times10^{-16}$ m$^4$V$^{-4}$~~~~~~& $7.75\times10^{-19}$ m$^4$V$^{-4}$~~~~~~ & ~~~photon-atom interaction\\
$\chi_{p2}^{(5)}$ ~~~~~~& $2.13\times10^{-12}$ m$^4$V$^{-4}$~~~~~~& $2.09\times10^{-14}$ m$^4$V$^{-4}$~~~~~~& Rydberg-Rydberg interaction\\
\hline
\end{tabular}
\end{table}
We now calculate the numerical values of the third-order and the fifth-order nonlinear optical susceptibilities
in the system based on the experimental parameters as given above. The other system parameters are selected as $n=60$, ${\cal N}_a=3\times10^{10}$\,cm$^{-3}$ and $\Delta_3=2\pi\times0.8\,{\rm MHz}$. By using the solutions presented in
Sec.~\ref{sec3} and the susceptibility formulas given in Eq.~(\ref{chi3}),  we obtain the third-order nonlinear optical susceptibilities $\chi_{p\alpha}^{(3)}={\rm Re}(\chi_{p\alpha}^{(3)})+i{\rm Im}(\chi_{p\alpha}^{(3)})$ ($\alpha=1,2$)
and the fifth-order nonlinear optical susceptibilities $\chi_{p\alpha}^{(5)}={\rm Re}(\chi_{p\alpha}^{(5)})+i{\rm Im}(\chi_{p\alpha}^{(5)})$ ($\alpha=1,2$) of the system, which are listed in Table I. Note that the value at the second row and the second column in the table is the real part of $\chi_{p1}^{(3)}$ (i.e. ${\rm
Re}(\chi_{p1}^{(3)})=-4.4\times10^{-11}\,{\rm m}^2\,{\rm V}^{-2})$, and the value at the second row and the third column is the imaginary part of $\chi_{p1}^{(3)}$ (i.e. ${\rm Im}(\chi_{p1}^{(3)})=-1.3\times10^{-13}\,{\rm m}^2 {\rm V}^{-2}$), etc. In the last column of Table I, the physical origins of various nonlinear optical susceptibilities of the system are given.

From Table I, we can obtain the following conclusions: (i)~The nonlinear optical susceptibilities in the present
Rydberg-EIT system are much larger than those obtained with conventional optical media such as optical fibers.
They are also larger than that obtained by using conventional EIT~\cite{Hau}.
(ii)~For the given atomic density (i.e. ${\cal N}_a=3\times10^{10}$\,cm$^{-3}$), the nonlinear optical susceptibilities
contributed by the Rydberg-Rydberg interaction (i.e. $\chi_{p2}^{(3)}$ and $\chi_{p2}^{(5)}$)
are three orders of magnitude greater than those contributed by the photon-atom interaction
(i.e. $\chi_{p1}^{(3)}$ and $\chi_{p1}^{(5)}$). Thus at this atomic density the Rydberg-Rydberg interaction
plays a leading role for the contribution of the nonlinear optical susceptibilities in the system.
In particular, the fifth-order nonlinear optical susceptibility originating from the Rydberg-Rydberg
interaction can reach the order of magnitude of $10^{-12}$\,m$^4$V$^{-4}$.
(iii)~The imaginary parts of the all nonlinear optical susceptibilities are much smaller than their corresponding real parts, which means that the nonlinear absorption can be suppressed in the nonlinear optical processes of the system. The physical reason for such suppression of the nonlinear absorption is the quantum interference effect induced by the control field (i.e. EIT effect) and also the introduction of the larger one-photon detuning $\Delta_2$, which makes the system have the giant optical nonlinearity of dispersive type~\cite{note00}.

The most interesting property of the nonlinear optical susceptibilities in the system is their dependence on the atomic density and the probe-field intensity when the other physical parameters are fixed~[23].
Fig.~\ref{fig2}(a) shows the real part of the third-order nonlinear optical susceptibilities
%
\begin{figure}\label{chi35}
\centering
\includegraphics[width=1\textwidth]{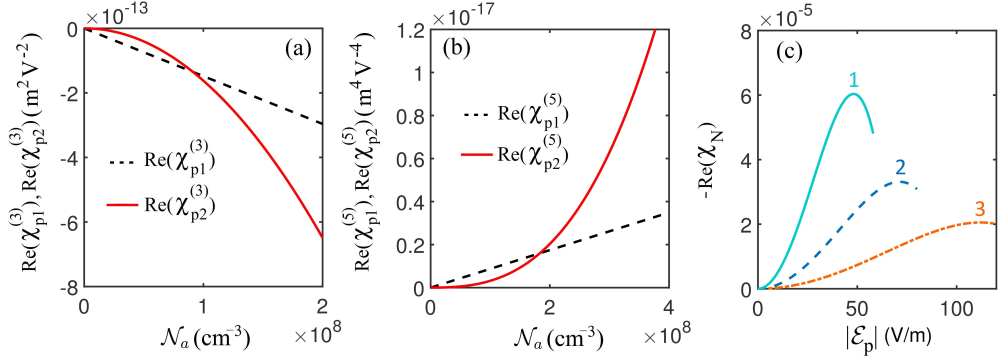}
\caption{\footnotesize(a) Re($\chi_{p1}^{(3)}$) (dashed-dotted line) and Re($\chi_{p2}^{(3)}$) (red solid line) as functions of  atomic density ${\cal N}_a$. (b) Re($\chi_{p1}^{(5)}$) (dashed-dotted line)  and Re($\chi_{p2}^{(5)}$) (red solid line) as functions of atomic density ${\cal N}_a$. (c) Total nonlinear optical susceptibility ${\rm Re}(\chi_N)={\rm Re}(\chi_p^{(3)}|\mathcal{E}_p|^2+\chi_p^{(5)}|\mathcal{E}_p|^4)$ as a function of $|{\cal E}_p|$. Lines from 1 to 3 correspond to ${\cal N}_a$= $4\times10^{10}$ cm$^{-3}$, $2\times10^{10}$ cm$^{-3}$ and $1\times10^{10}$ cm$^{-3}$, respectively.}
\label{fig2}
\end{figure}
Re($\chi_{p1}^{(3)}$) (black dashed-dotted line) and Re($\chi_{p2}^{(3)}$) (red solid line) as functions of ${\cal N}_a$. When plotting the figure, the parameters used are the same as those used for getting the results in the Table I except for ${\cal N}_a$, which is now taken as a variable. From the figure we see that: (i)~The Kerr nonlinearities contributed by the photon-atom interaction and the Rydberg-Rydberg interaction are comparable with atomic density ${\cal N}_a$ around $10^8$\,cm$^{-3}$, and both of them are increasing functions of ${\cal N}_a$. (ii)~For ${\cal N}_a$ less than $0.9\times 10^8$\,cm$^{-3}$, the Kerr nonlinearity by the Rydberg-Rydberg interaction is smaller than that by the photon-atom interaction. However, the both Kerr nonlinearities arrive at the same value ${\rm Re}(\chi_{p1}^{(3)})={\rm Re}(\chi_{p2}^{(3)})=-1.3 \times 10^{-13}\,{\rm m}^2{\rm V}^{-2}$ at ${\cal N}_a=0.9\times 10^8$\,cm$^{-3}$. (iii)~When ${\cal N}_a$ is larger than $0.9\times 10^8$\,cm$^{-3}$ the Kerr nonlinearity by the Rydberg-Rydberg interaction surpasses that by the photon-atom interaction and grows rapidly as ${\cal N}_a$ increases.

Shown in Fig.~\ref{fig2}(b) are the real parts of the fifth-order nonlinear optical susceptibilities
Re($\chi_{p1}^{(5)}$) (black dashed-dotted line) and Re($\chi_{p2}^{(5)}$) (red solid line) as functions of ${\cal N}_a$, which originate respectively from the photon-atom and the Rydberg-Rydberg interactions.
We observe that both Re($\chi_{p1}^{(5)}$) and Re($\chi_{p2}^{(5)}$)
are comparable, and they are decreasing functions of ${\cal N}_a$. Furthermore, for ${\cal N}_a$ less than $1.86\times 10^8$\,cm$^{-3}$, the fifth-order nonlinear optical susceptibility by the Rydberg-Rydberg interaction is larger than that by the photon-atom interaction. At ${\cal N}_a=1.86\times 10^8$\,cm$^{-3}$, the both optical susceptibilities become equal to have the value ${\rm Re}(\chi_{p1}^{(5)})={\rm Re}(\chi_{p2}^{(5)})=1.67 \times 10^{-18}\,{\rm m}^4{\rm V}^{-4}$. When ${\cal N}_a$ is larger than $1.86\times 10^8$\,cm$^{-3}$ ${\rm Re}(\chi_{p2}^{(5)})$ becomes to be smaller than ${\rm Re}(\chi_{p1}^{(5)})$.

From the above results, we see that there exist various, synergetic optical nonlinearities in the Rydberg-EIT system.
Due to the ${\cal N}_a^2$- and ${\cal N}_a^3$-dependence, the nonlinear optical susceptibilities contributed by the Rydberg-Rydberg interaction are sensitive to the atomic density and hence they can exceed the optical nonlinearities contributed by the photon-atom interaction for large ${\cal N}_a$.  From Fig.~2(a) and Fig.~2(b) we also see that the sign of $\chi_{p\alpha}^{(3)}$ is opposite to that of $\chi_{p\alpha}^{(5)}$ ($\alpha=1,2$) and $\chi_{p\alpha}^{(5)}$ grows faster than $\chi_{p\alpha}^{(3)}$, which means that there exists a competition between $\chi_{p\alpha}^{(3)}$ and $\chi_{p\alpha}^{(5)}$ when ${\cal N}_a$ becomes larger.

Different from Fig.~\ref{fig2}(a) and Fig.~\ref{fig2}(b), where the dependence of the third-order and the fifth-order nonlinear optical susceptibilities on the atomic density ${\cal N}_a$ are illustrated, in
Fig.~2(c) we show ${\rm Re}(\chi_N)={\rm Re}(\chi_p^{(3)})|\mathcal{E}_p|^2+{\rm Re}(\chi_p^{(5)})|\mathcal{E}_p|^4$, i.e. the real part of the total nonlinear optical susceptibility of the probe field, as a function of $|\mathcal{E}_p|$ ($\mathcal{E}_p$ is the envelope of the probe field) for several different atomic density. Lines from 1 to 3 in the figure are for ${\cal N}_a$ taken to be $6\times10^{10}$ cm$^{-3}$, $4\times10^{10}$ cm$^{-3}$, and $3\times10^{10}$ cm$^{-3}$, respectively. We observe that, for all ${\cal N}_a$,
Re$(\chi_N)$ grows fast initially, then arrives a peak value, and  finally decreases as $|\mathcal{E}_p|$ increases.
We also observe that the higher the atomic density, the faster Re$(\chi_N)$ arrives to its peak value,
which is due to the effect coming from the fifth-order nonlinear susceptibilities.
We stress that the maximum value of the probe field used in Fig.~\ref{fig2}(c) is
$|{\cal E}_{p{\rm max}}|=120$ V$/$m, which is within the validity domain of the perturbation theory used above
because $|\Omega_{p{\rm max}}/\Omega_c|\simeq0.1$.

For comparison, in Fig.~\ref{fig3}(a) (Fig.~\ref{fig3}(b)\,)
%
\begin{figure}
\centering
\includegraphics[width=0.8\textwidth]{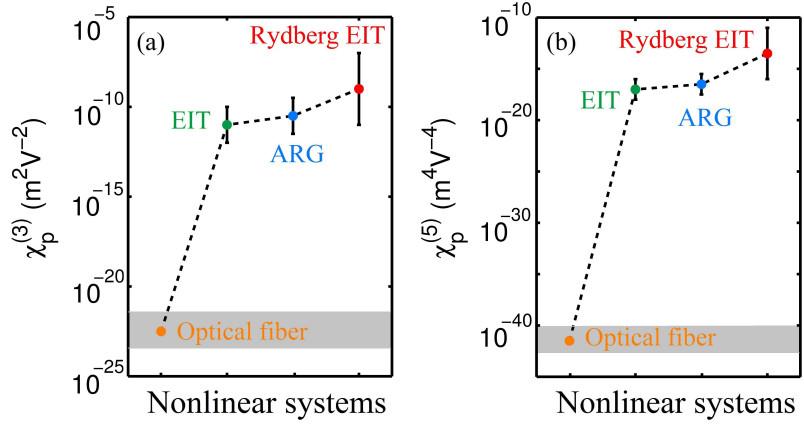}
\caption{\footnotesize(a) ((b)\,) Third-order (Fifth-order) nonlinear optical susceptibility $\chi_p^{(3)}$ ($\chi_p^{(5)}$) for optical fibers~\cite{Agrawal} (yellow solid circle), conventional EIT~\cite{Hau} (green solid circle), ARG system~\cite{ARG,ARG1} (blue solid circle), and the present Rydberg-EIT system (red solid circle),  respectively. The black vertical line at each solid circle indicates the range of the nonlinear optical susceptibility for the atomic density varying from $10^{9}$\,cm$^{-3}$ to $6\times10^{10}$\,cm$^{-3}$. The Grey shaded area symbolizes the range of the nonlinear optical susceptibilities for optical fibers.}\label{fig3}
\end{figure}
%
we show the third-order (fifth-order) nonlinear optical susceptibility obtained for several typical physical systems, including optical fibers~\cite{Agrawal}, conventional EIT system~\cite{Hau}, active Raman gain (ARG) system~\cite{ARG,ARG1}, and the present Rydberg EIT system, with the value of $\chi_p^{(3)}$ ($\chi_p^{(5)}$) indicated by the yellow solid circle, green solid circle, blue solid circle, and red solid circle, respectively. The black vertical line at each solid circle indicates the range of the nonlinear optical susceptibility for the atomic density varying from $10^{9}$\,cm$^{-3}$ to $6\times10^{10}$\,cm$^{-3}$. The Grey shaded area in the lower part of the figure symbolizes the range of the nonlinear optical susceptibilities for optical fibers. We see that the third-order and the fifth-order nonlinear optical susceptibilities
obtained by using the present Rydberg EIT system have the highest values in comparison with the other systems~\cite{Hau,Agrawal,ARG,ARG1}.
Especially, the third-order nonlinear optical susceptibilities in the present Rydberg EIT system can reach the order of magnitude of $10^{-7}\,{\rm m}^2{\rm V}^{-2}$ for atomic density ${\cal N}_a=6\times10^{10}$\,cm$^{-3}$, which agrees fairly with reported experimental and theoretical results~\cite{note2}. If the atomic density increases to ${\cal N}_a=5.0\times 10^{12}$ cm$^{-3}$ (used in \cite{Hau}), we obtain
$\chi_p^{(3)}=4.6\times 10^{-3}$ m$^2$V$^{-2}$. Thus the third-order nonlinear optical susceptibility in the present Rydberg-EIT system
is five orders of magnitude larger than that obtained in the conventional EIT system, where
$\chi_p^{(3)}=7\times 10^{-8}$ m$^2$V$^{-2}$~\cite{Hau}.
Furthermore, The fifth-order nonlinear optical susceptibilities in the present Rydberg EIT system can reach the order of magnitude of $10^{-11}\,{\rm m}^4{\rm V}^{-4}$, which is five orders of magnitude larger than that of the conventional EIT
system and the ARG systems for the peak atomic density ${\cal N}_a=6\times 10^{10}$\,cm$^{-3}$.
The physical reasons for such giant third-order and fifth-order optical Kerr effects obtained in the present Rydberg-EIT system are the cooperative response of a large number of atoms, the quantum interference contribution from the EIT, and the Rydberg-Rydberg interaction in the system.  Such giant third-order and the fifth-order optical nonlinearities are very promising for the investigation of many nonlinear optical processes not possible by using conventional optical media up to now.

Note that when taking $\Delta_2\ll\Gamma_{12}$, we can also gain another type of optical nonlinearity of
the system. For instance, if we choose ${\cal N}_a=3\times10^{10}$\,cm$^{-3}$, $\Omega_c=2\pi\times16\,\,${\rm MHz}, $\Delta_2=1\,\,{\rm kHz}$ and $\Delta_3=0$ (the blockade sphere radius $R_b=3.26\,{\rm \mu m}$) and the other parameters the same as those given above we obtain $\chi_p^{(3)}=(-7.07+i3.4)\times10^{-10}\,{\rm m}^2\,{\rm V}^{-2}$ and $\chi_p^{(5)}=(2.03+0.8)\times10^{-14}\,{\rm m}^4\,{\rm V}^{-4}$. In this situation, imaginary parts of $\chi_p^{(3)}$ and $\chi_p^{(5)}$ have the same orders of magnitude as their corresponding real parts, i.e. the system displays an optical nonlinearity of dissipative type. The large imaginary part in $\chi_p^{(3)}$ and
$\chi_p^{(5)}$ will result inevitably in high photon loss for the nonlinear behavior in the system. Notice that the dissipative-type third-order nonlinear optical susceptibility was considered in \cite{Sevincli}, in which the photon-atom interaction has negligible contribution to optical susceptibilities. The reasons are the following. (i)~The Rydberg state has a long lifetime (i.e. $\Gamma_{23}$ is very small); (ii)~The two-photon detuning $\Delta_3$ was taken to be zero. As a result, in \cite{Sevincli} only the Rydberg-Rydberg interaction contributes to the nonlinear optical susceptibilities. In our work, both the third-order and the fifth-order nonlinear optical susceptibilities have been considered by taking a non-zero $\Delta_3$, and hence both the photon-atom interaction and the Rydberg-Rydberg interaction play significant roles for the optical nonlinearity of the system.

\section{Summary}\label{sec6}

In this article, we have investigated the optical Kerr effects in an ensemble of cold Rydberg atoms via EIT. By using an approach beyond mean-field approximation, we have proved that the system can possess not only an enhanced third-order nonlinear optical susceptibility, which has a ${\cal N}_a^2$-dependence, but also a giant fifth-order nonlinear optical susceptibility, which has ${\cal N}_a^2$- and ${\cal N}_a^3$-dependence. We have demonstrated that both the third-order and the fifth-order nonlinear optical susceptibilities consist of two parts, which are contributed respectively by the photon-atom interaction and the strong Rydberg-Rydberg interaction. The Kerr nonlinearity induced by the Rydberg-Rydberg  interaction plays a leading role for high atomic density. We have found that the fifth-order nonlinear susceptibility in the present Rydberg-EIT system may be five orders of magnitude larger than that obtained in traditional EIT systems, which may have promising applications in light and quantum information processing and transmission at few photon level. The theoretical method proposed here is systematic and can be used to calculate other high-order (e.g. the seventh-order, ninth-order, etc.) nonlinear optical susceptibilities~\cite{note03}, and for other Rydberg states (e.g. $nD$ states for which the dipole-dipole interaction is angular dependent~\cite{Ravets,Tresp}). It can be also generalized to investigate non-instantaneous optical Kerr effects in cold, interacting Rydberg systems.

\section*{Appendix}

\appendix

\section{Expansion of the one-body density matrix equation }\label{ap2}
Under the perturbation expansion $\rho_{\alpha1}=\Omega_p\sum_{l=0}\rho_{\alpha1}^{(2l+1)}|\Omega_p|^{2l}$, $\rho_{32}=\sum_{l=1}\rho_{32}^{(2l)}|\Omega_p|^{2l}$, $\rho_{\beta\beta}=\sum_{l=0}\rho_{\beta\beta}^{(2l)}|\Omega_p|^{2l}$
with $\rho_{\beta\beta}^{(0)} =\delta_{\beta1}\delta_{\beta1}$($\alpha=2,3; \beta=1,2,3$), the one-body density matrix equation (\ref{eq2}) becomes
\begin{subequations} \label{Exp1}
\begin{eqnarray}
&&\left(i\frac{\partial }{\partial t}+d_{21}\right)
\rho_{21}^{(l)}+1+\Omega_{c}^{\ast}\rho_{31}^{(l)}=A^{(l)},\\
&&\left(i\frac{\partial }{\partial t}+d_{31}\right)\rho_{31}^{(l)}+\Omega_{c}\rho_{21}^{(l)}=B^{(l)},\\
&&i\frac{\partial }{\partial t}\rho_{11}^{(l)}+i\Gamma_{12}\left(\rho_{11}^{(l)}+\rho_{33}^{(l)}\right)=\rho_{12}^{(l-1)}-\rho_{21}^{(l-1)},\\
&&i\frac{\partial }{\partial t}\rho_{33}^{(l)}+i\Gamma_{23}\rho_{33}^{(l)}+\Omega_{c}\rho_{23}^{(l)}-\Omega_{c}^{\ast }\rho_{32}^{(l)}=0,\\
&&\left(i\frac{\partial }{\partial t}+d_{32}\right) \rho_{32}^{(l)}-\Omega_{c}(2\rho_{33}^{(l)}+\rho_{11}^{(l)})=C^{(l)}.
\end{eqnarray}
\end{subequations}
Here $A^{(1)}=A^{(2)}=B^{(1)}=B^{(2)}=C^{(1)}=0$,
$A^{(l)}=-2\rho_{11}^{(l-1)}+\rho_{33}^{(l-1)}$ ($l=3,4,5$), $B^{(l)}=-\rho_{32}^{(l-1)}+{\cal
N}_a\int{d^3\mathbf{r}^\prime V(\mathbf{r}^\prime-\mathbf{r})\rho\rho_{33,31}^{(l)}(\mathbf{r}^\prime,\mathbf{r},t)}$ ($l=3,4,5$), $C^{(2)}=\rho_{31}^{(1)}$, $C^{(3)}=\rho_{31}^{(2)}$, and $C^{(l)}=\rho_{31}^{(l-1)}+{\cal
N}_a\int{d^3\mathbf{r}^\prime V(\mathbf{r}^\prime-\mathbf{r})\rho\rho_{33,32}^{(l)}(\mathbf{r}^\prime,\mathbf{r},t)}$ ($l=4,5$).

\section{Expansion of the equations of the two-body correlators}\label{ap3}

By a simple inspection on the order of magnitude for the two-body density matrix (correlator) elements
$\langle \hat{S}_{\alpha\beta}\hat{S}_{\mu\nu}\rangle\equiv \rho\rho_{\alpha\beta,\mu\nu}$
based on the weak driven EIT condition (i.e. the probe-field is weak), we have the expansion $\rho\rho_{\alpha\beta,\mu\nu}=\rho\rho_{\alpha\beta,\mu\nu}^{(2)}\Omega_p^2+\rho\rho_{\alpha\beta,\mu\nu}^{(4)}\Omega_p^2|\Omega_p|^2+\cdots$.
By a detailed and careful calculation, we obtain the following equations of motion for the two-body correlators
from second-order to fifth-order approximations:

(i) {\it Second-order approximation} ($l=2$). For the two-body correlators, the lowest-order approximation starts from $\epsilon^2$-order. We obtain the equations
\begin{align}\label{twobody1}
&\begin{bmatrix}2d_{21} & 0 & 2\Omega_c^\ast \\
    0 & 2d_{31}-V & 2\Omega_c \\
    \Omega_c & \Omega_c^\ast & d_{21}+d_{31}
\end{bmatrix}
\begin{bmatrix}
\rho\rho_{21,21}^{(2)}\\ \rho\rho_{31,31}^{(2)}\\ \rho\rho_{31,21}^{(2)}
\end{bmatrix}\nonumber\\
&=\begin{bmatrix}
-2\frac{d_{31}}{D}\\0\\ \frac{\Omega_c}{D}
\end{bmatrix},
\end{align}

and
\begin{align}\label{twobody2}
&\begin{bmatrix}d_{21}+d_{12} & 0 & -\Omega_c & \Omega_c^\ast\\
    -\Omega_c^\ast & \Omega_c^\ast & d_{21}+d_{13} & 0 \\
    0 & d_{31}+d_{13} & \Omega_c &-\Omega_c^\ast \\
    -\Omega_c & \Omega_c &0 & d_{21}^\ast+d_{13}^\ast
\end{bmatrix}
\begin{bmatrix}
\rho\rho_{21,12}^{(2)}\\ \rho\rho_{31,13}^{(2)}\\ \rho\rho_{21,13}^{(2)} \\ \rho\rho_{21,13}^{\ast(2)}
\end{bmatrix}\nonumber\\
&=\begin{bmatrix}
\frac{d_{31}}{D}-\frac{d_{31}^\ast}{D^\ast}\\ \frac{\Omega_c^\ast}{D^\ast}\\0\\ \frac{\Omega_c}{D}
\end{bmatrix},
\end{align}

(ii) {\it Third-order approximation} ($l=3$). At this order, we have the equations
\begin{align}\label{C4}
  &\begin{bmatrix}\begin{matrix}M_{31} & \Omega_c^\ast & -i\Gamma_{23} & 0 & \Omega_c^\ast & -\Omega_c & 0 & 0 \\
    \Omega_c & M_{32} & 0 & -i\Gamma_{23} & 0 & 0 & \Omega_c^\ast & -\Omega_c \\
    0 & 0 & M_{33} & \Omega_c^\ast & -\Omega_c^\ast & \Omega_c & 0 & 0 \\
    0 & 0 & \Omega_c & M_{34} & 0 & 0 & -\Omega_c^\ast & \Omega_c \\
    \Omega_c & 0 & -\Omega_c & 0 & M_{35} & 0 & \Omega_c^\ast & 0 \\
    -\Omega_c^\ast & 0 & \Omega_c^\ast & 0 & 0 & M_{36} & 0 & \Omega_c^\ast \\
    0 & \Omega_c & 0 & -\Omega_c & \Omega_c & 0 & M_{37} & 0 \\
    0 & -\Omega_c^\ast & 0 & \Omega_c^\ast & 0 & \Omega_c & 0 & M_{38}
\end{matrix}\end{bmatrix}\begin{bmatrix}
\begin{matrix}\rho\rho_{22,21}^{(3)}\\ \rho\rho_{22,31}^{(3)}\\ \rho\rho_{33,21}^{(3)}\\ \rho\rho_{33,31}^{(3)}\\ \rho\rho_{32,21}^{(3)}\\ \rho\rho_{21,23}^{(3)}\\ \rho\rho_{32,31}^{(3)}\\ \rho\rho_{31,23}^{(3)}
\end{matrix}
\end{bmatrix}\nonumber\\
&=\begin{bmatrix}
\begin{matrix}-\rho\rho_{21,12}^{(2)}+\rho\rho_{21,21}^{(2)}-\rho_{22}^{(2)}\\
    -\rho\rho_{31,12}^{(2)}+\rho\rho_{21,31}^{(2)}\\-\rho_{33}^{(2)}\\0\\ \rho\rho_{21,31}^{(2)}-\rho_{32}^{(2)}\\-\rho_{32}^{\ast(2)}-\rho\rho_{21,13}^{(2)}\\ \rho\rho_{31,31}^{(2)}\\
    -\rho\rho_{31,13}^{(2)}
\end{matrix}
\end{bmatrix},
\end{align}
where $M_{31}=i\Gamma_{12}+d_{21}$, $M_{32}=i\Gamma_{12}+d_{31}$, $M_{33}=i\Gamma_{23}+d_{21}$, $M_{34}=d_{31}+i\Gamma_{23}-V$, $M_{35}=d_{32}+d_{21}$, $M_{36}=d_{23}+d_{21}$, $M_{37}=d_{32}+d_{31}-V$ and $M_{38}=d_{23}+d_{31}$. The general expression of $\rho\rho_{33,31}^{(3)}$ reads
\begin{equation}\label{order33313}
\rho\rho_{33,31}^{(3)}=\frac{P_0+P_1V(\mathbf{r}^\prime-\mathbf{r})+P_2V(\mathbf{r}^\prime-\mathbf{r})^2}{Q_0
+Q_1V(\mathbf{r}^\prime-\mathbf{r})+Q_2V(\mathbf{r}^\prime-\mathbf{r})^2+Q_3V(\mathbf{r}^\prime-\mathbf{r})^3},
\end{equation}
Here $P_n$ and $Q_n~(n=0,1,2,3)$ are functions of the spontaneous emission decay rate $\gamma_{\mu\nu}$, detunings $\Delta_\mu$ and half Rabi frequency $\Omega_c$. The third-order nonlinear susceptibility $\chi_{p2}^{(3)}$ can be obtained by integrating $\rho\rho_{33,31}^{(3)}$ (see
Eq.~(\ref{thirdorder})\,).

(iii) {\it Fourth-order approximation} ($l=4$). At this order, one has the equations
\begin{align}\label{C5}
&\begin{bmatrix}\begin{matrix}i\Gamma_{12} & 0 & -i\Gamma_{23} & \Omega_c^\ast & 0 & 0 & 0 & -\Omega_c & 0 & 0 \\
    0 & i\Gamma_{23} & 0 & 0 & 0 & 0 & -\Omega_c^\ast & 0 & \Omega_c & 0\\
    0 & -i\Gamma_{23} & M_{43} & -\Omega_c^\ast & 0 & 0 & \Omega_c^\ast & \Omega_c & -\Omega_c & 0 \\
    \Omega_c & 0 & -\Omega_c & M_{44} & \Omega_c^\ast & -\Omega_c & -i\Gamma_{23} & 0 & 0 & 0\\
    0 & 0 & 0 & 2\Omega_c & M_{45} & 0 & -2\Omega_c & 0 & 0 & 0\\
    0 & 0 & 0 & -\Omega_c^\ast & 0 & M_{46} & \Omega_c^\ast & \Omega_c & -\Omega_c & 0\\
    0 & -\Omega_c & \Omega_c & 0 & -\Omega_c^\ast & \Omega_c & M_{47} & 0 & 0 & 0\\
    -\Omega_c^\ast & 0 & \Omega_c^\ast & 0 & 0 & \Omega_c^\ast & 0 & M_{48} & -i\Gamma_{23} & -\Omega_c\\
    0 & \Omega_c^\ast & -\Omega_c^\ast & 0 & 0 & -\Omega_c^\ast & 0 & 0 & M_{49} & \Omega_c\\
    0 & 0 & 0 & 0 & 0 & 0 & 0 & -2\Omega_c^\ast & 2\Omega_c^\ast & M_{40}
\end{matrix}\end{bmatrix}\begin{bmatrix}
\begin{matrix}\rho\rho_{22,22}^{(4)}\\ \rho\rho_{33,33}^{(4)}\\ \rho\rho_{33,22}^{(4)}\\ \rho\rho_{22,32}^{(4)}\\ \rho\rho_{32,32}^{(4)}\\ \rho\rho_{23,32}^{(4)}\\
    \rho\rho_{33,32}^{(4)}\\ \rho\rho_{22,23}^{(4)}\\ \rho\rho_{33,23}^{(4)}\\ \rho\rho_{23,23}^{(4)}
\end{matrix}
\end{bmatrix}\nonumber\\
&=\begin{bmatrix}
\begin{matrix}\rho\rho_{22,21}^{(3)}-\rho\rho_{22,21}^{\ast(3)}\\0\\
    \rho\rho_{33,21}^{(3)}-\rho\rho_{33,21}^{\ast(3)}\\ \rho\rho_{32,21}^{(3)}+\rho\rho_{22,31}^{(3)}-\rho\rho_{21,23}^{\ast(3)}\\2\rho\rho_{32,31}^{(3)}\\ \rho\rho_{31,23}^{(3)}-\rho\rho_{32,13}^{(3)}\\ \rho\rho_{33,31}^{(3)}\\ \rho\rho_{21,23}^{(3)}-\rho\rho_{32,21}^{\ast(3)}-\rho\rho_{22,31}^{\ast(3)}\\-\rho\rho_{33,31}^{\ast(3)}\\
    -2\rho\rho_{32,31}^{\ast(3)}
\end{matrix}
\end{bmatrix},
\end{align}
where $M_{43}=i\Gamma_{12}+i\Gamma_{23}$, $M_{44}=d_{32}+i\Gamma_{12}$, $M_{45}=2d_{32}-V$, $M_{46}=d_{32}+d_{23}$, $M_{47}=i\Gamma_{23}+d_{32}-V$, $M_{48}=d_{23}+i\Gamma_{12}$, $M_{49}=i\Gamma_{23}+d_{23}+V$, $M_{40}=2d_{23}+V$. A general expression for the radial dependence of $\rho\rho_{33,32}^{(4)}$ is given by
\begin{equation}\label{aa33324}
\rho\rho_{33,32}^{(4)}=\frac{\sum_{n=0}^6K_nV(\mathbf{r}^\prime-\mathbf{r})^n}
{\sum_{n=0}^7J_nV(\mathbf{r}^\prime-\mathbf{r})^n},
\end{equation}
where $K_n$ and $J_n$ are functions of the spontaneous emission decay rate $\gamma_{\mu\nu}$, detunings $\Delta_\mu$ and half Rabi frequency $\Omega_c$. The fourth order of atomic population $\rho_{\alpha\alpha}^{(4)} (\alpha=1, 2, 3)$ can be calculated by integrating $\rho\rho_{33,32}^{(4)}$.

Another part of equations at the fourth-order reads
\begin{align}\label{C6}
&\begin{bmatrix}2d_{21} & 0 & 2\Omega_c^\ast \\
    0 & 2d_{31}-V & 2\Omega_c \\
    \Omega_c & \Omega_c^\ast & d_{21}+d_{31}
\end{bmatrix}
\begin{bmatrix}
\rho\rho_{21,21}^{(4)}\\ \rho\rho_{31,31}^{(4)}\\ \rho\rho_{31,21}^{(4)}
\end{bmatrix}\nonumber\\
&=\begin{bmatrix}
4\rho\rho_{22,21}^{(3)}+2\rho\rho_{33,21}^{(3)}-2a_{21}^{(3)}\\2\rho\rho_{32,31}^{(3)}\\ 2\rho\rho_{22,31}^{(3)}+\rho\rho_{33,31}^{(3)}+\rho\rho_{32,21}^{(3)}-a_{31}^{(3)}
\end{bmatrix}\nonumber\\
&+N_a\begin{bmatrix}
-2b_{21}^{(3)}\\ 2\int d^3\mathbf{r}^{\prime\prime} \left(\frac{-2\Omega_c \rho\rho_{33,31}^{(3)}}{D}+\rho_{33}^{(2)}\rho\rho_{31,31}^{(2)}-\frac{\Omega_c^2 \rho_{33}^{(2)}}{D^2}\right)V(\mathbf{r}^{\prime\prime}-\mathbf{r})\\ -b_{31}^{(3)}+\int d^3\mathbf{r}^{\prime\prime} \left(\frac{d_{31}\rho\rho_{33,31}^{(3)}}{D}-\frac{\Omega_c \rho\rho_{33,21}^{(3)}}{D}+\rho_{33}^{(2)}\rho\rho_{31,21}^{(2)}+\frac{\Omega_cd_{31}\rho_{33}^{(2)}}{D^2}\right)V(\mathbf{r}^{\prime\prime}-\mathbf{r})
\end{bmatrix},
\end{align}

and
\begin{align}\label{C7}
&\begin{bmatrix}d_{21}+d_{12} & 0 & -\Omega_c & \Omega_c^\ast\\
     0 & d_{31}+d_{13} & \Omega_c &-\Omega_c^\ast \\
    -\Omega_c^\ast & \Omega_c^\ast & d_{21}+d_{13} & 0 \\
    -\Omega_c & \Omega_c &0 & d_{21}^\ast+d_{13}^\ast
\end{bmatrix}
\begin{bmatrix}
\rho\rho_{21,12}^{(4)}\\ \rho\rho_{31,13}^{(4)}\\ \rho\rho_{21,13}^{(4)} \\ \rho\rho_{21,13}^{\ast(4)}
\end{bmatrix}\nonumber\\
=&\begin{bmatrix}
2(\rho\rho_{22,21}^{\ast(3)}-\rho\rho_{22,21}^{(3)})+\rho\rho_{33,21}^{\ast(3)}-\rho\rho_{33,21}^{(3)}+b_{21}^{(3)}-b_{21}^{\ast(3)}\\ \rho\rho_{31,23}^{\ast(3)}-\rho\rho_{31,23}^{(3)}\\
2\rho\rho_{22,31}^{\ast(3)}+\rho\rho_{33,31}^{\ast(3)}-a_{31}^{\ast(3)}-\rho\rho_{21,23}^{(3)}\\ 2\rho\rho_{22,31}^{(3)}+\rho\rho_{33,31}^{(3)}-a_{31}^{(3)}-\rho\rho_{21,23}^{\ast(3)}
\end{bmatrix}\nonumber\\
&+{\cal N}_a\begin{bmatrix}
b_{21}^{(3)}-b_{21}^{\ast(3)}\\ 0\\ -b_{31}^{\ast(3)}+\int d^3\mathbf{r}^{\prime\prime} \left(\frac{d_{31}\rho\rho_{33,31}^{\ast(3)}}{D}-\frac{\Omega_c^\ast\rho\rho_{33,21}^{(3)}}{D^\ast}
+\rho_{33}^{(2)}\rho\rho_{21,13}^{(2)}+\frac{\rho_{33}^{(2)}\Omega_c^\ast d_{31}}{|D|^2}\right)
V(\mathbf{r}^{\prime\prime}-\mathbf{r}) \\ -b_{31}^{(3)}+\int d^3\mathbf{r}^{\prime\prime} \left(\frac{d_{31}^\ast\rho\rho_{33,31}^{(3)}}{D^\ast}-\frac{\Omega_c\rho\rho_{33,21}^{\ast(3)}
}{D}+\rho_{33}^{(2)}\rho\rho_{21,13}^{\ast(2)}+\frac{\rho_{33}^{(2)}\Omega_c
d_{31}^\ast}{|D|^2}\right)V(\mathbf{r}^{\prime\prime}-\mathbf{r})
\end{bmatrix},
\end{align}
The solution of the Eqs.~(\ref{C6}) and (\ref{C7}) is given by $\rho\rho_{\alpha1,\beta1}^{(4)}=aa_{\alpha1,\beta1}^{(4)}+{\cal N}_abb_{\alpha1,\beta1}^{(4)}$, and $\rho\rho_{\alpha1,1\beta}^{(4)}=aa_{\alpha1,1\beta}^{(4)}+{\cal N}_abb_{\alpha1,1\beta}^{(4)}(\alpha,\beta=2,3)$. The explicit expressions of $aa_{\alpha1,\beta1}^{(4)}$, $aa_{\alpha1,1\beta}^{(4)}$,
$bb_{\alpha1,\beta1}^{(4)}$ and $bb_{\alpha1,1\beta}^{(4)}$ can be easily obtained by using Cramer$^\prime$s rule, which are lengthy and omitted here for saving space.

(ii) {\it Fifth-order approximation} ($l=5$). An this order, we have the equations
\begin{align}\label{C8}
  &\begin{bmatrix}\begin{matrix}M_{31} & \Omega_c^\ast & -i\Gamma_{23} & 0 & \Omega_c^\ast & -\Omega_c & 0 & 0 \\
    \Omega_c & M_{32} & 0 & -i\Gamma_{23} & 0 & 0 & \Omega_c^\ast & -\Omega_c \\
    0 & 0 & M_{33} & \Omega_c^\ast & -\Omega_c^\ast & \Omega_c & 0 & 0 \\
    0 & 0 & \Omega_c & M_{34} & 0 & 0 & -\Omega_c^\ast & \Omega_c \\
    \Omega_c & 0 & -\Omega_c & 0 & M_{35} & 0 & \Omega_c^\ast & 0 \\
    -\Omega_c^\ast & 0 & \Omega_c^\ast & 0 & 0 & M_{36} & 0 & \Omega_c^\ast \\
    0 & \Omega_c & 0 & -\Omega_c & \Omega_c & 0 & M_{37} & 0 \\
    0 & -\Omega_c^\ast & 0 & \Omega_c^\ast & 0 & \Omega_c & 0 & M_{38}
\end{matrix}\end{bmatrix}\begin{bmatrix}
\begin{matrix}\rho\rho_{22,21}^{(5)}\\ \rho\rho_{22,31}^{(5)}\\ \rho\rho_{33,21}^{(5)}\\ \rho\rho_{33,31}^{(5)}\\ \rho\rho_{32,21}^{(5)}\\ \rho\rho_{21,23}^{(5)}\\ \rho\rho_{32,31}^{(5)}\\ \rho\rho_{31,23}^{(5)}
\end{matrix}
\end{bmatrix}\nonumber\\
=&\begin{bmatrix}
\begin{matrix}2\rho\rho_{22,22}^{(4)}+\rho\rho_{33,22}^{(4)}-aa_{21,12}^{(4)}+aa_{21,21}^{(4)}-a_{22}^{(4)}\nonumber\\
    aa_{31,21}^{(4)}-aa_{31,12}^{(4)}+\rho\rho_{22,32}^{(4)}\\
    2\rho\rho_{22,33}^{(4)}+\rho\rho_{33,33}^{(4)}-a_{33}^{(4)}\\
    \rho\rho_{33,32}^{(4)}\\
    2\rho\rho_{22,32}^{(4)}+\rho\rho_{33,32}^{(4)}+aa_{31,21}^{(4)}-a_{32}^{(4)}\\
    2\rho\rho_{22,32}^{\ast(4)}+\rho\rho_{33,32}^{\ast(4)}-aa_{21,13}^{(4)}-a_{32}^{\ast(4)}\\
    \rho\rho_{32,32}^{(4)}+aa_{31,31}^{(4)}\\
    \rho\rho_{32,23}^{(4)}-aa_{31,13}^{(4)}
\end{matrix}
\end{bmatrix}\nonumber\\
&+{\cal N}_a\begin{bmatrix}
\begin{matrix}-b_{22}^{(4)}-bb_{21,12}^{(4)}+bb_{21,21}^{(4)}\\
    bb_{31,21}^{(4)}-bb_{31,12}^{(4)}+N_{12}\\
    -b_{33}^{(4)}\\
    N_{14}\\
    bb_{31,21}^{(4)}-b_{32}^{(4)}+N_{15}\\
    -b_{32}^{\ast(4)}-bb_{21,13}^{(4)}-N_{16}\\
    bb_{31,31}^{(4)}+N_{17}\\
    -bb_{31,13}^{(4)}
\end{matrix}
\end{bmatrix},
\end{align}\label{Fif}
where $N_{12}=\int d^3\mathbf{r}^{\prime\prime} (\rho\rho_{33,31}^{(3)}\rho_{22}^{(2)}+\rho\rho_{22,21}^{(3)}\rho_{33}^{(2)}-\rho\rho_{22,33}^{(4)}
\Omega_c/D+\rho_{33}^{(2)}\rho_{22}^{(2)}\Omega_c/D)V(\mathbf{r}^{\prime\prime}-\mathbf{r})$,
$N_{14}=\int d^3\mathbf{r}^{\prime\prime} (2\rho\rho_{33,31}^{(3)}\rho_{33}^{(2)}-\rho\rho_{33,33}^{(4)}\Omega_c/D+\rho_{33}^{(2)}\rho_{33}^{(2)}
\Omega_c/D)V(\mathbf{r}^{\prime\prime}-\mathbf{r})$, $N_{15}=\int d^3\mathbf{r}^{\prime\prime} (\rho\rho_{32,21}^{(3)}\rho_{33}^{(2)}+\rho_{33,21}^{(3)}\rho_{32}^{(2)}
+\rho\rho_{33,32}^{(4)}d_{31}/D-\rho_{33}^{(2)}\rho_{32}^{(2)}d_{31}/D)
V(\mathbf{r}^{\prime\prime}-\mathbf{r})$, $N_{16}=\int d^3\mathbf{r}^{\prime\prime} (\rho\rho_{23,21}^{(3)}\rho_{33}^{(2)}+\rho\rho_{33,21}^{(3)}\rho_{32}^{\ast(2)}+\rho\rho_{33,32}^{\ast(4)}
d_{31}/D-\rho_{33}^{(2)}\rho_{32}^{\ast(2)}d_{31}/D)V(\mathbf{r}^{\prime\prime}-\mathbf{r})$, $N_{17}=\int d^3\mathbf{r}^{\prime\prime} [\rho\rho_{33,31}^{(3)}(\rho_{32}^{(2)}+\rho_{22}^{(2)})+(\rho\rho_{32,31}^{(3)}+\rho\rho_{22,31}^{(3)})\rho_{33}^{(2)}-(\rho\rho_{33,32}^{(4)}+\rho\rho_{33,22}^{(4)})\Omega_c/D+\rho_{33}^{(2)}(\rho_{32}^{(2)}+\rho_{22}^{(2)})\Omega_c/D]V(\mathbf{r}^{\prime\prime}-\mathbf{r})$.

The solution of Eq.~(\ref{C8}) is given by $\rho\rho_{\alpha\alpha,\beta1}^{(5)}=aa_{\alpha\alpha,\beta1}^{(5)}+{\cal N}_abb_{\alpha\alpha,\beta1}^{(5)}$, $\rho\rho_{32,\alpha1}^{(5)}=aa_{32,\alpha1}^{(5)}+{\cal N}_abb_{32,\alpha1}^{(5)}$ and $\rho\rho_{\alpha1,23}^{(5)}=aa_{\alpha1,23}^{(5)}+{\cal N}_abb_{\alpha1,23}^{(5)}$$(\alpha,\beta=2,3)$. General expressions for the radial dependence of $aa_{33,31}^{(5)}$ and $bb_{33,31}^{(5)}$ are given as:
\begin{eqnarray}\label{order5}
&&aa_{33,31}^{(5)}=\frac{\sum_{n=0}^8W_nV(\mathbf{r}^\prime-\mathbf{r})^n}
{\sum_{n=0}^{9}Y_nV(\mathbf{r}^\prime-\mathbf{r})^n}, \label{aa33315}\\
&&bb_{33,31}^{(5)}=\frac{X_0+X_1V(\mathbf{r}^\prime-\mathbf{r})}{Z_0+Z_1
V(\mathbf{r}^\prime-\mathbf{r})+Z_2V(\mathbf{r}^\prime-\mathbf{r})^2}, \label{bb33315}
\end{eqnarray}
were $W_n$, $Y_n$, $X_n$ and $Z_n$ are functions of the spontaneous emission decay rate $\gamma_{\mu\nu}$, detunings $\Delta_\mu$ and half Rabi frequency $\Omega_c$. The fifth-order nonlinear susceptibility $\chi_{p2}^{(5)}$ can be obtained by integrating these general forms analytically. The explicit expressions of $aa_{\alpha\alpha,\beta1}^{(5)}$, $aa_{32,\alpha1}^{(5)}$, $aa_{\alpha1,23}^{(5)}$, $bb_{\alpha\alpha,\beta1}^{(5)}$, $bb_{32,\alpha1}^{(5)}$ and $bb_{\alpha1,23}^{(5)}$ are omitted here.

\section*{Acknowledgments}

This work was supported by the NSF-China under Grants No.~11174080 and No.~11474099, and by the Chinese Education Ministry Reward for Excellent Doctors in Academics under Grant No.~PY2014009.


\begin{thebibliography}{99}
\bibitem{note000}Because $R_b$ is  independent of atomic density ${\mathcal N}_a$, the increase of ${\mathcal N}_a$ results in only an increase of atoms (which are not in the Rydberg state) in the blockade sphere.
\bibitem{note111}Researches~\cite{pritchard2,stan,Sche} showed that if the atomic density ${\cal{N}}_a$ is large than $4\times10^{8}$cm$^{-3}$ the mean-field
approximation is broken down.
\bibitem{note2}The third-order nonlinear optical susceptibilities in Rydberg atomic gases can arrive
at the order of magnitude of $10^{-7}\,{\rm m}^2{\rm V}^{-2}$ and has a cubic dependence on atomic density, which have been reported in experment (see Refs.~\cite{Pri,pritchard2,pritchard1,Parigi}) and theory (see Refs.~\cite{Sevincli,Petrosyan,stan}).
\bibitem{note03} Since these high-order nonlinear optical susceptibilities are much smaller than the third- and
fifth-order ones, the discussion on them is omitted here.
\end{thebibliography}
\end{document}